\documentclass[a4paper,fleqn,usenatbib]{mnras}

\usepackage{newtxtext,newtxmath}
\usepackage{mathptmx}

\usepackage[T1]{fontenc}
\usepackage{ae,aecompl}

\usepackage{graphicx}	
\usepackage{amsmath}	
\usepackage{amssymb}	

\title[Two radio-weak BL Lac candidates with VLBI]{The loud and the quiet: searching for radio counterparts of two radio-weak BL Lac candidates with VLBI}

\author[H.-M. Cao et al.]{Hong-Min Cao,$^{1,2}$\thanks{E-mail: \href{mailto:hongmin.cao@foxmail.com}{hongmin.cao@foxmail.com} (H-MC)}
S\'{a}ndor Frey,$^{3}$
Krisztina \'{E}. Gab\'{a}nyi,$^{3,4}$
Jun Yang,$^{5}$
Lang Cui,$^{1}$ 
\newauthor
Xiao-Yu Hong$^{6}$
and Tao An$^{6}$
\\
$^{1}$Xinjiang Astronomical Observatory, Chinese Academy of Sciences, 150 Science 1-Street, Urumqi, Xinjiang 830011, China\\
$^{2}$School of Electronic and Electrical Engineering, Shangqiu Normal University, Wenhua Road 298, Shangqiu, Henan 476000, China\\
$^{3}$Konkoly Observatory, MTA Research Centre for Astronomy and Earth Sciences, Konkoly Thege Mikl\'{o}s \'{u}t 15-17, H-1121 Budapest, Hungary\\
$^{4}$MTA-ELTE Extragalactic Astrophysics Research Group, P\'{a}zm\'{a}ny P\'{e}ter s\'{e}t\'{a}ny 1/A, H-1117 Budapest, Hungary\\
$^{5}$Department of Space, Earth and Environment, Chalmers University of Technology, Onsala Space Observatory, SE-43992 Onsala, Sweden\\
$^{6}$Shanghai Astronomical Observatory, Key Laboratory of Radio Astronomy, Chinese Academy of Sciences, 80 Nandan Road, Shanghai 200030,\\
China
}

\date{Accepted 2018 XXX NN. Received 2018 XXX NN; in original form 2018 July 26}

\pubyear{2018}

\begin{document}
\label{firstpage}
\pagerange{\pageref{firstpage}--\pageref{lastpage}}
\maketitle

\begin{abstract}

BL Lac objects are known to have compact jets inclined to our line of sight at a small angle, showing prominent radio emission. Two radio-weak BL Lac candidates with no counterparts in current radio surveys were recently reported by \citet{Massaro2017}. Both sources were selected as candidate low-energy counterparts of unassociated {\em Fermi} $\gamma$-ray sources. We carried out very long baseline interferometry (VLBI) observations with the European VLBI Network (EVN) at 5~GHz to explore their radio properties at milli-arcsecond (mas) scale. One target, J1410+7405, is clearly detected with the EVN. Its measured 5-GHz flux density, 2.4~mJy, is consistent with recent interferometric measurements with the Karl G. Jansky Very Large Array, suggesting that the radio emission is confined to the inner $\la 10$~mas region. J1410+7405 is therefore identified as a radio-loud jetted active galactic nucleus, and its brightness temperature exceeds $\sim 10^{9}$~K. Its properties are similar to those of other $\gamma$-ray detected BL Lac objects. On the other hand, the second target, J0644+6031, remains undetected with the EVN with a 6-$\sigma$ brightness upper limit of 0.12~mJy~beam$^{-1}$. This source is thus radio-quiet, confirming its peculiarity, or possibly questioning its BL Lac nature.

\end{abstract}




\section{Introduction}
\label{sec-1}

Blazars, a sub-class of active galactic nuclei (AGN), produce prominent non-thermal emission across the entire electromagnetic spectrum. This emission could even dominate over the thermal components coming from the accretion disk, the obscuring torus and the host galaxy. Blazars typically exhibit flux density variability on multiple time scales, from minutes to years, and often show polarization in radio and optical. All these observing characteristics are ascribed to the twin relativistic jets closely aligned with the line of sight, launched from the AGN central engine, an accreting supermassive black hole. The approaching jet contributes the most due to the Doppler-boosting effect \citep[see][for a review]{Urry1995}. 

Blazars are traditionally divided into two main subclasses: BL Lac objects and flat-spectrum radio quasars (FSRQs). The former have no or weak optical emission lines, while the latter have pronounced emission lines. The rest-frame equivalent width (REW) of 5~\AA\,\,is usually considered as the dividing line between these two source populations \citep[e.g.][]{Stickel1991}. The broad-band $\log \nu  F_{\nu} - \log \nu$ spectral energy distributions (SEDs) of blazars show a double-hump feature which can be generally explained in the context of the one-zone leptonic model \citep{Ghisellini2009}. Blazars are the dominant emitters in the $\gamma$-ray sky. Among the identified/associated {\em Fermi} sources in the third Large Area Telescope (LAT) source catalogue (3FGL), about 57 per cent are blazars \citep{Acero2015}. 

In the {\em Wide-field Infrared Survey Explorer (WISE)} mid-infrared (mid-IR) color--color diagram, blazars appear to be located in a region different from those occupied by thermal sources. This is the {\em WISE} blazar strip (WBS), where the thermal emission is contaminated by the non-thermal emission produced by the relativistic jet \citep{Massaro2011}. The region of the $\gamma$-ray blazars is even narrower, and is called the {\em WISE} gamma-ray strip \citep[WGS,][]{D'Abrusco2012,Massaro2012}. The {\em WISE} mid-IR colors falling into the WGS are widely considered as a piece of evidence for classifying a source as a blazar.

Very recently, \citet{Massaro2017} identified two sources, WISE J064459.38+603131.7 (hereafter J0644+6031) and WISE J141046.00+740511.2 (hereafter J1410+7405), as genuine radio-weak BL Lacs (RWBLs)\footnote{Here the term ``radio-weak'' means no radio counterparts known in the current major radio surveys, thus low radio flux density. However, \citet{Bruni2018} adopt 1.4-GHz monochromatic radio power $L_{1.4} < 10^{25.5}$\,W\,Hz$^{-1}$, suggested by \citet{Gregg1996} for radio-quiet AGN, to define RWBLs.}. They are particularly interesting because they have no radio counterparts at 1.4~GHz in the U.S. National Radio Astronomy Observatory (NRAO) Very Large Array (VLA) Sky Survey\footnote{\url{https://www.cv.nrao.edu/nvss/}} \citep[NVSS,][]{Condon1998}. Both sources are outside the sky coverage of the VLA Faint Images of the Radio Sky at Twenty-centimeters\footnote{\url{http://sundog.stsci.edu/}} survey \citep[FIRST,][]{White1997}. Radio emission is usually seen as a representative characteristic of blazars with relativistic jets. It is thus puzzling why these two alleged BL Lacs remained undetected in radio. Are they really silent, or do they have radio emission which is unknown just because it is below the detection limit ($\sim 2.5$~mJy) of the NVSS?

We observed J0644+6031 and J1410+7405 using the technique of very long baseline interferometry (VLBI) with the European VLBI Network (EVN) at 5~GHz. These VLBI observations were sufficiently sensitive to detect compact emission if the sources are radio-loud. VLBI can provide direct proof of the blazar nature of an AGN by means of detecting milli-arcsecond (mas) scale radio emission originating from the inner jet. The aim of our EVN experiment was thus confirming or ruling out that these two intriguing objects are BL Lacs.

The observed characteristics of the two targets are introduced in Section~\ref{sec-2}. The details of the VLBI experiment and the data reduction are described in Section~\ref{sec-3}. The results are presented in Section~\ref{sec-4} and discussed in Section~\ref{sec-5}. A summary is given in Section~\ref{sec-6}. In this letter, we use a flat $\Lambda$CDM model with $H_0 = 71$\,km\,s$^{-1}$\,Mpc$^{-1}$, $\Omega_{\rm m} = 0.27$ and $\Omega_{\Lambda} = 0.73$ \citep{Spergel2007} for cosmological distance estimation.

\section{Target sources}
\label{sec-2}

J0644+6031 was selected as a blazar candidate counterpart of 2FGL~J0644.6+6034 \citep{Massaro2013}, an unassociated $\gamma$-ray source (UGS) in the second {\em Fermi} LAT source catalogue \citep[2FGL,][]{Nolan2012}. The nearest USNO-B \citep{Monet2003} optical counterpart to this {\em WISE} source is just $0\farcs3$ away\footnote{\url{http://vizier.u-strasbg.fr/viz-bin/VizieR-3?-source=I/284}}. Follow-up optical spectroscopic observations by \citet{Paggi2014} revealed weak emission lines with REWs consistent with the threshold of 5~\AA, supporting the BL Lac classification, and suggest a redshift of $z = 0.3582 \pm 0.0008$.

J1410+7405 is coincident with one of the two X-ray sources found in the position error ellipse of an UGS 1FHL J1410.4+7408 \citep{Landi2015}, listed in the first {\em Fermi} LAT high-energy source catalogue \citep[1FHL,][]{Ackermann2013}. Only one USNO-B optical source, $0\farcs4$ away from J1410+7405, is located in the error circle of the X-ray source \citep[$\sim 6\arcsec$ in radius,][]{Landi2015}. It has an optical spectrum typical of BL Lacs, barely showing any emission or absorption lines and therefore prohibiting redshift determination \citep{Marchesini2016}.

Both {\em WISE} sources (J0644+6031 and J1410+7405) lie in the subregion on the WGS populated by the $\gamma$-ray BL Lacs. They also show flux density variation known from their light curves built from the {\em WISE} single-epoch photometric observations covering $\sim 2$ days each. Moreover, the multi-band photometric data of these two sources can be well described by a log-parabolic function often used to fit the non-thermal SEDs of BL Lacs. The properties of their optical spectra, mid-IR colors and variability, and the shape of their broad-band SEDs together suggest the blazar nature of these two {\em WISE} sources \citep{Massaro2017}. 

\begin{table*}
	\centering
	\caption{EVN observing information.}
	\label{tab-1}
	\begin{tabular}{lccccccc}
		\hline
		\hline
		Project segment & Target & Right ascension & RA err. & Declination & Dec err. & Phase calibrator & Separation \\
		                &        & h m s           & mas     & \degr \,\, \arcmin \,\, \arcsec & mas &   & \degr      \\
		\hline
		EC061A & J1410+7405 & 14 10 45.9505 & 4.5 & +74 05 10.905 & 4.1 & J1353+7532 & 1.85 \\
		EC061B & J0644+6031 & 06 44 59.3851 & 1.4 & +60 31 31.655 & 1.6 & J0650+6001 & 0.85 \\
		\hline
	\end{tabular}
\\
{\bf Notes:} Cols.~3--6 -- a priori equatorial coordinates (J2000) assumed for pointing to the target sources and their errors, taken from the {\em Gaia} DR1 catalogue \citep{Gaia2016a,Gaia2016b}; Col.~8 -- angular separation between the target and the phase calibrator. 
\end{table*}

\section{Observations and data reduction}
\label{sec-3}

\subsection{Observations}

The 5-GHz EVN experiment, divided into two segments with project codes EC061A and EC061B, was conducted on 2017 October 26. There were thirteen participating antennas in both segments: Effelsberg (Ef, Germany), Jodrell Bank Mk2 (Jb, United Kingdom), Medicina (Mc, Italy), Toru\'n (Tr, Poland), Westerbork (Wb, the Netherlands), Noto (Nt, Italy), Yebes (Ys, Spain), Tianma (T6, China), Urumqi (Ur, China), Svetloe (Sv, Russia), Badary (Bd, Russia), Zelenchukskaya (Zc, Russia), and Irbene (Ir, Latvia). For most of the antennas, the data were recorded at 2048~Mbps rate, with 2 polarizations, 8 intermediate frequency channels (IFs) per polarization, and 32~MHz bandwidth per IF. For Jb, Tr, and Wb, 4 IFs per polarization were used, with a recording rate of 1024~Mbps. The duration of each segment was about 2~h. The observations were carried out in phase-referencing mode \citep[e.g.][]{Beasley1995}, i.e., the antennas nodded between the nearby bright and compact phase calibrator and the target. The cycle time was 5-min long, with an on-target time of 3.5~min in each cycle. Short 5-min scans of bright fringe-finder sources (J1419+5423 and DA193 for segments EC061A and EC061B, respectively) were also scheduled. The recorded data from the individual stations were correlated at the Joint Institute for VLBI ERIC (JIVE) in Dwingeloo, the Netherlands, with 2~s integration time and 64 spectral channels per IF. Further observing information is given in Table~\ref{tab-1}. In this experiment, the {\em Gaia} Data Release 1 (DR1) coordinates\footnote{\url{https://gea.esac.esa.int/archive/}} \citep{Gaia2016a,Gaia2016b} of the optical counterparts of the target sources were used for pointing the radio telescopes and for the correlation. 

\subsection{Data reduction}

The VLBI data were calibrated in the NRAO Astronomical Image Processing System \citep[{\sc aips},][]{Greisen2003} in both segments separately, according to the EVN Data Reduction Guide\footnote{\url{http://www.evlbi.org/user\_guide/analysis.html}}. A priori amplitude calibration was carried out using the known antenna gain curves and system temperatures regularly measured at the VLBI stations during the observations, or the nominal system equivalent flux densities where the system temperatures were not available. Parallactic angle and ionospheric corrections were then performed successively. Manual phase calibration, global fringe-fitting and bandpass calibration were also done. 

The calibrated visibility data of the phase-reference and fringe-finder sources were exported to the {\sc difmap} program \citep{Shepherd1997} for hybrid mapping. The antenna-based amplitude correction factors obtained by the {\sc gscale} command were fed back into {\sc aips} using the task {\sc clcor}, to adjust the visibility amplitudes for the target sources as well. The factors lower than $\pm$5 per cent were not applied but corrections were typically higher. Fringe-fitting was then repeated in {\sc aips} for the phase-reference calibrators, but this time their brightness distribution models produced in {\sc difmap} were taken into account. This eliminates the influence of the structure that may cause residual phase errors. Finally, the solutions obtained for the reference sources were interpolated to the targets in {\sc aips}, and the phase-referenced calibrated visibility data were analysed in {\sc difmap}.

\section{Results}
\label{sec-4}

\begin{figure}
	\centering
	\includegraphics[width=0.9\columnwidth]{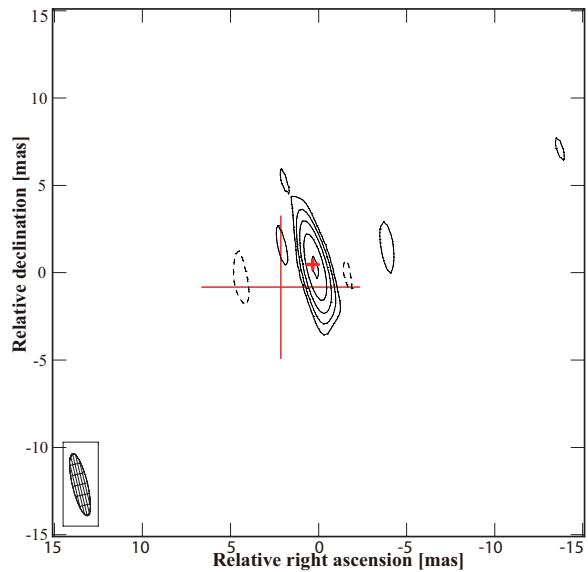}
    \caption{Phase-referenced VLBI image of J1410+7405 at 5~GHz, made with natural weighting and centred on the peak position. The first contours are drawn at $\pm 129$\,$\mu$Jy\,beam$^{-1}$ ($\pm5\sigma$ image noise level). The positive contours increase by a factor of 2. The peak brightness is 2.2~mJy\,beam$^{-1}$. The Gaussian restoring beam with 3.6\,mas~$\times$~0.9\,mas (FWHM) is shown in the lower left corner. The position angle of the beam major axis is 13$\degr$, measured from north through east. The crosses denote the source optical positions taken from {\em Gaia} DR1 (Table \ref{tab-1}) and {\em Gaia} DR2 (see the text), with their sizes indicating the uncertainties.}
    \label{fig-1}
\end{figure}

{\bf J1410+7405} is clearly detected with VLBI. A single circular Gaussian model component was used to fit the source brightness distribution in the visibility domain in {\sc difmap}. Due to the weakness of the source, self-calibration was not attempted. The naturally-weighted modelfit image is shown in Fig.~\ref{fig-1}. VLBI phase-referencing allows us to measure the accurate J2000 coordinates of the radio brightness peak in J1410+7405. The right ascension ${\rm RA} = 14^{\rm h} 10^{\rm m} 45\fs95002$ and declination ${\rm DEC} = +74\degr 05\arcmin 10\farcs9061$ were determined from the image using the {\sc aips} task {\sc tvmaxfit}. The estimated positional uncertainty of $\sim$ 0.2~mas is dominated by the systematic error components, i.e., the positional uncertainty of the phase-reference calibrator J1353+7532, and the uncertainty caused by the target--calibrator angular separation \citep{Pradel2006}. The statistical error is negligible in this case, due to the high dynamic range of the phase-referenced image. The {\em Gaia} DR1 optical position (see Table~\ref{tab-1} and Fig.~\ref{fig-1}) is consistent with that measured with VLBI within the errors, which strengthens the physical association of the optical and radio source. Note that the most recent {\em Gaia} DR2 \citep{Gaia2018} optical position ${\rm RA} = 14^{\rm h} 10^{\rm m} 45\fs95006$ and ${\rm DEC} = 74\degr 05\arcmin 10\farcs9063$ (uncertainty $\sim$ 0.3~mas, Fig. \ref{fig-1}) is also consistent with and is even closer to our VLBI position.

The measured parameters of J1410+7405 are listed in Table~\ref{tab-2}. The brightness temperature $T_{\rm b}$ in the source rest frame is calculated as
\begin{equation}
    T_{\rm b}=1.22 \times 10^{12}(1+z)\frac{S_{\nu}}{\theta^2\nu_{\rm obs}^2}\,\,{\rm K}
	\label{eq-1}
\end{equation}
\citep{Condon1982}, where $z$ is the redshift, $S_{\nu}$ the observed flux density in Jy, $\theta$ the Gaussian component size (full width at half maximum, FWHM) in mas, and $\nu_{\rm obs}$ the observing frequency in GHz. Since the redshift of J1410+7405 is unknown, we derived a lower limit ($T_{\rm b} > 1.2 \times 10^9$~K) assuming $z = 0$. This value suggests that the radio emission originates from a jet and is produced by non-thermal synchrotron radiation. The fitted component size ($0.28 \pm 0.01$~mas, Table~\ref{tab-2}) is slightly smaller than the minimum resolvable size \citep[e.g.][]{Kovalev2005}, 0.31~mas, indicating the source is unresolved. The latter value was used to calculate the brightness temperature. The strong phase coherence on one of the longest and the most sensitive Ef--T6 baseline confirms the compactness of J1410+7405.  

In case of our other target, {\bf J0644+6031}, we searched for compact radio emission in the naturally weighted dirty map with a field of view as large as $2\arcsec \times 2\arcsec$, even though the a priori {\em Gaia} position was accurate to mas level (Table \ref{tab-1}). However, nothing was detected above a signal-to-noise ratio of 6. The 6-$\sigma$ brightness upper limit is 120~$\mu$Jy\,beam$^{-1}$. We conclude that any mas-scale compact radio source in J0644+6031 should be weaker than 0.12~mJy (Table \ref{tab-2}). 

We repeated the analysis to constrain the VLBI network to relatively shorter baselines by removing data from T6, Ur and Bd, to become more sensitive to $\sim 10$~mas scale extended radio emission. For J1410+7405, one circular Gaussian component fitting gave a flux density of $2.5 \pm 0.5$ mJy, consistently with that measured with the full VLBI array (Table \ref{tab-2}), which is another support for the source compactness. However, no radio source was found for J0644+6031. 

\begin{table}
	\centering
	\caption{Parameters of the targets derived from the 5-GHz EVN observations.}
	\label{tab-2}
	\begin{tabular}{lccc} 
		\hline
		\hline
		Source name & $S_{\nu}$    & $\theta$ & $T_{\rm b}$ \\
		            &   mJy        & mas      & 10$^9$ K    \\
		\hline
		J1410+7405  & 2.4 $\pm$ 0.4   & 0.28 $\pm$ 0.01 & $>1.2$ \\
		J0644+6031  & $<0.12$         & -               & -      \\
		\hline
	\end{tabular}
\\
{\bf Notes:} Col.~2 -- flux density; Col.~3 -- circular Gaussian model component diameter (FWHM); Col.~4 -- brightness temperature. Modelfit errors are estimated according to \citet{Lee2016} based on the 3-$\sigma$ image noise level. For the flux density, additional 10 per cent error was considered, to account for the VLBI amplitude calibration uncertainty.
\end{table}

\section{Discussion}
\label{sec-5}

Among the {\em Fermi} $\gamma$-ray sources, a large fraction still have no known counterparts in other electromagnetic wavebands. For example, in 3FGL, about 30 per cent remain to be identified/associated \citep{Acero2015}. Finding their low-energy counterparts is essential to reveal the physical origin of the $\gamma$-ray emission, and thus to better understand the whole $\gamma$-ray sky. Considering that a large percentage of the associated {\em Fermi} sources are blazars, if there is a blazar discovered within the positional error ellipse (normally at 95 per cent confidence level) of an UGS, then this blazar would be the most likely object where the $\gamma$-rays come from. 

The compact radio emission detected with VLBI proves the existence of jet activity in J1410+7405. This mid-IR source is positionally coincident with UGS 1FHL J1410.4+7408 \citep[a.k.a. UGS 3FGL J1410.9+7406,][]{Landi2015,Massaro2017}. Based on the multi-band data available so far, J1410+7405 is the most promising $\gamma$-ray emitter in the {\em Fermi} error ellipse.

This source is also detected with the Karl G. Jansky VLA at 5~GHz ($2.2 \pm 0.1$~mJy) and 7~GHz ($2.0 \pm 0.1$~mJy), in a large observing project carried out very recently, aiming to search for radio counterparts of all UGSs in 3FGL \citep{Schinzel2017}. The deduced flat spectrum with spectral index $\alpha = -0.28 \pm 0.28$ is consistent with a compact synchrotron self-absorbed radio source (the spectral index is defined as $S_{\nu} \varpropto \nu^{\alpha}$). Our 5-GHz VLBI flux density ($2.4 \pm 0.4$~mJy) is identical within the errors with that measured by the VLA, suggesting that the radio emission essentially comes from a small mas-scale region probed by the EVN. In addition, the accurate match of the source positions measured with VLBI and {\em Gaia}, and the high compactness (the ratio of the VLBI and VLA flux densities close to 1) imply that the jet is oriented at a small viewing angle \citep[e.g.][]{Liuzzo2013}. However, the VLA experiment \citep{Schinzel2017} was carried out more than 2 years before our EVN observations. We therefore cannot rule out flux density variability in J1410+7405 which is typical for compact blazars.

The relative strength of AGN jet activity could be characterised with the radio-loudness parameter $R = S_{\rm r}/S_{\rm o}$. Here, $S_{\rm r}$ and $S_{\rm o}$ denote the observed flux densities at 5~GHz and 4400\,\AA, respectively \citep{Kellermann1989}. Alternatively, the rest-frame flux density ($S_{\nu,\rm {rest}} \varpropto \frac{\nu^\alpha}{(1+z)^{1+\alpha}}$) is also used to calculate the radio loudness \citep[e.g.][]{Wu2013}. As the redshift (for J1410+7405) and the optical spectral indices (for both targets) are not known, we used the classical definition of radio loudness \citep{Kellermann1989} with observed flux densities, but this will not change our conclusions. AGN are traditionally classified as radio-quiet if $R<10$ and radio-loud if $R \ge 10$. Known from the USNO-A2.0 catalogue\footnote{\url{http://vizier.u-strasbg.fr/viz-bin/VizieR-3?-source=I/252}} \citep{Monet1998}, the optical B-band magnitude of J1410+7405 is 19.2. This, together with the 5-GHz radio flux density, gives $R \approx 40$, indicating that J1410+7405 is radio-loud.

Taking the median redshift value of 1FHL BL Lac objects, i.e., $z = 0.2$ \citep{Lico2016}, we estimate the 1.4-GHz monochromatic radio power $L_{1.4} = (3.1 \pm 0.3) \times 10^{23}$ W\,Hz$^{-1}$ for J1410+7405. This value is compatible with the RWBL definition adopted by \citet{Bruni2018}. However, if the source is at $z \ga 1.7$, its power $L_{1.4} \ge 10^{25.5}$ W\,Hz$^{-1}$ would classify it as radio-loud according to \citet{Gregg1996}, consistently with our radio-loudness estimation by the \citet{Kellermann1989} criterion.

The jet Doppler factor $\delta$ can be inferred from a single-epoch VLBI observation, i.e. $\delta = T_{\rm {b}}/T_{\rm {eq}}$, where $T_{\rm {eq}} \approx 5 \times 10^{10}$~K is the equipartition brightness temperature \citep{Readhead1994}. A Doppler factor larger than unity would indicate Doppler-boosted emission and is often used as an important criterion for classifying a source as a blazar \citep[e.g.][]{Coppejans2016}. The measured brightness temperature of J1410+7405 ($T_{\rm b} > 1.2 \times 10^9$ K) is only a lower limit and thus the Doppler factor is $\delta \ga 0.02$. 
Even if a non-zero redshift is assumed in Eq.~\ref{eq-1}, the low limit of the core brightness temperature does not usefully constrain the Doppler factor in our case, leaving the question open whether the jet emission in J1410+7405 is not boosted or simply the angular resolution is insufficient to measure $\delta \ga 1$. The core can be blended with a more extended jet component, resulting in an overestimated angular size and thus an underestimation of the source brightness temperature \citep{Natarajan2017}.

The $\gamma$-ray blazars in the {\em Roma-BZCAT}\footnote{\url{http://www.asdc.asi.it/bzcat/}} \citep{Massaro2015} show a bimodal feature in the distribution of the logarithmic flux density ratio between 1.4~GHz in the radio and 3.4~$\mu$m in mid-IR \citep[see Fig.~10 in][]{Massaro2017}. The available radio data of J1410+7405 allow us to calculate ${\rm log}(S_{\rm 1.4 GHz}/S_{\rm 3.4 \mu m}) = 1.0$, which is different from the value below 0.5 as expected by \citet{Massaro2017}, and places the source to the region where $\gamma$-ray BL Lacs populate the distribution.

Our second VLBI target, J0644+6031, is positionally consistent with UGS 2FGL J0644.6+6034, but lies outside the 95 per cent confidence positional error ellipse of UGS 3FGL J0644.6+6035, which has a somewhat more accurate position \citep[see Fig.~8 in][]{Massaro2017}. Thus its physical association with the $\gamma$-ray source is rather uncertain. However, not all known blazars appear detected by {\em Fermi} LAT \citep{Lister2015}. Our 5-GHz flux density upper limit (120 $\mu$Jy; Table \ref{tab-2}) for the mas-scale compact radio emission and the optical B-band magnitude \citep[USNO(B)=19,][]{Monet1998} provide a stringent upper limit of $R \la 1.5$ for the radio loudness of J0644+6031. Assuming spectral index $\alpha = 0$, we obtain $L_{1.4} < 3.8 \times 10^{22}$ W\,Hz$^{-1}$ and ${\rm log}(S_{\rm 1.4 GHz}/S_{\rm 3.4 \mu m}) < -0.86$ for J0644+6031. The radio non-detection itself and the radio quietness confirm the unusual characteristics of J0644+6031 as a BL Lac object suggested by \citet{Massaro2017}. From our EVN data, there is no evidence for a radio jet in this source. However, it is also possible that the BL Lac identification of J0644+6031 is questionable. In this case, the BL Lac-type characteristics in wavebands other than the radio would call for alternative explanations. Indeed, J0644+6031 has a noisier optical spectrum than J1410+7405 \citep{Paggi2014,Marchesini2016}. A spectrum like this could be caused by the intrinsically weak broad emission lines, rather than those swamped by the continuum emission from the relativistic jet \citep{Plotkin2010}. Based on the data currently available, the nature of J0644+6031 remains elusive, and further efforts are required to draw a final conclusion.

\begin{figure*}
	\centering
	\includegraphics[width=2.0\columnwidth,trim = 1 1 1 1,clip]{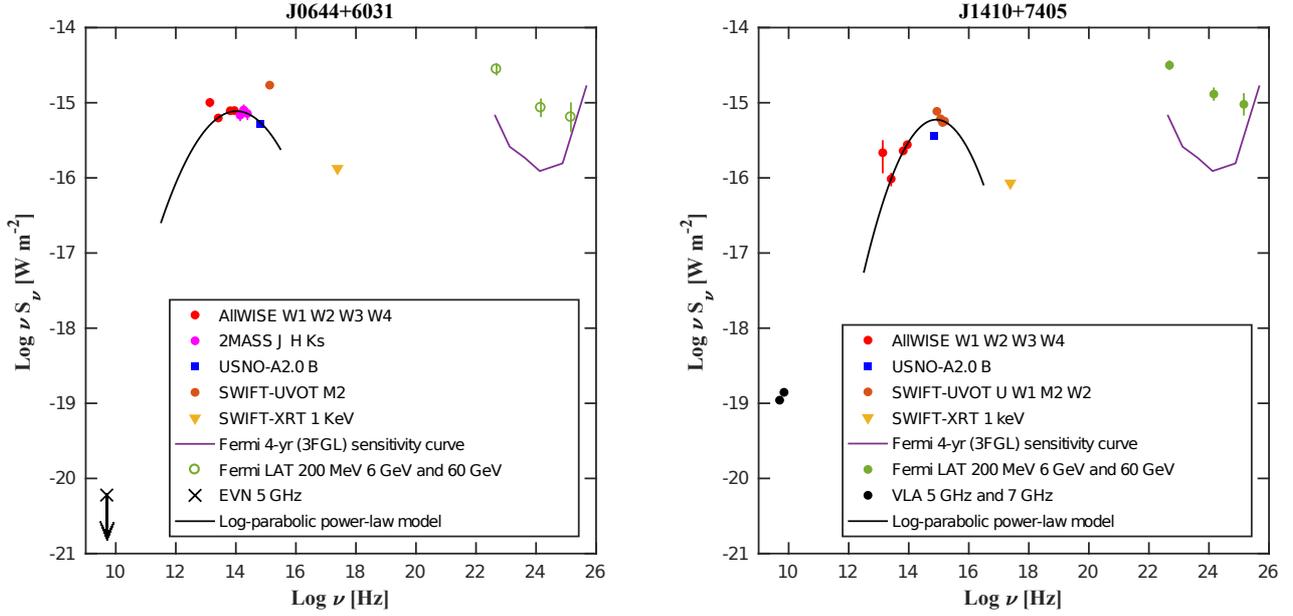}
    \caption{The broadband SEDs of the two EVN targets J0644+6031 (left panel) and J1410+7405 (right panel). As J0644+6031 is not detected with the EVN, the 5-GHz flux upper limit (marked with $\times$) is shown in the left panel. Three $\gamma$-ray data points are plotted to illustrate the {\em Fermi} detection. Since the association of J0644+6031 with the {\em Fermi} $\gamma$-ray source is uncertain, being outside the 95 per cent confidence positional error ellipse of 3FGL J0644.6+6035 \citep{Massaro2017}, the $\gamma$-ray data points are marked with open circles. The {\em Fermi} 4-yr sensitivity curves, based on 3FGL, correspond to a Test Statistic (TS) of 2.}
    \label{fig-2}
\end{figure*}

The broadband SEDs of the radio source J1410+7405 and the radio-undetected J0644+6031 are shown in Fig.~\ref{fig-2}. We calculated the {\em WISE}, 2MASS, and {\em Swift} UVOT/XRT fluxes following \citet{Massaro2017}. The USNO-A2.0 B magnitude was first transformed to the Landolt magnitude\footnote{\url{http://www.quasars.org/docs/USNO_Landolt.htm}} and then corrected with the Galactic extinction according to \citet{Schlafly2011}. The $\gamma$-ray fluxes and the {\em Fermi} 4-year sensitivity curves were obtained with the web-based tool provided by the ASI Science Data Center (ASDC)\footnote{\url{https://tools.ssdc.asi.it/}}. The synchrotron peak frequencies derived from the log-parabolic fit (see Fig.~\ref{fig-2}) are $(8.2 \pm 2.2) \times 10^{14}$ Hz and $(1.0 \pm 0.3) \times 10^{14}$ Hz for J1410+7405 and J0644+6031, respectively. Thus, regarding the blazar classification by \citet{Abdo2010}, the former source should belong to the intermediate-synchrotron peaked (ISP) BL Lacs, while J0644+6031, if indeed a BL Lac, would be around the dividing line between the ISP and low-synchrotron peaked (LSP) objects.

\section{Summary and conclusions}
\label{sec-6}

We observed with the EVN two radio-weak BL Lac candidates, proposed as counterparts of unidentified {\em Fermi} $\gamma$-ray sources \citep{Massaro2017}. Our 5-GHz observations capable of mas-scale angular resolution did not detect any compact radio source at the position of J0644+6031. The inferred small value of radio-loudness parameter $R \la 1.5$ confirms the peculiarity of this object found by \citet{Massaro2017}. It might either be a true RWBL or its BL Lac identification should be revised since this class of objects is known to show synchrotron radio emission from relativistic jets closely aligned to the line of sight. Both the BL Lac nature of J0644+6031 and its association with a {\em Fermi} source would require further supporting evidence. 

On the other hand, compact radio emission from the source J1410+7405 is detected with VLBI. The measured brightness temperature of the 2.4-mJy component is in the order of $10^9$~K, not sufficient to prove that the jet emission is Doppler-boosted. Nevertheless, its properties are similar to those of other known 1FHL BL Lac objects. The radio loudness of the source is $R \approx 40$. The case of J1410+7405 demonstrates that the lack of detection in a relatively shallow radio survey like the NVSS does not necessarily preclude the existence of mas-scale compact radio emission in a radio-loud (jetted) AGN.

\section*{Acknowledgements}

We thank the anonymous referee for valuable suggestions.
The European VLBI Network is a joint facility of independent European, African, Asian, and North American radio astronomy institutes. Scientific results from data presented in this publication are derived from the following EVN project code: EC061. H-MC acknowledges support by the CAS ``Light of West China'' programme (Grants No. 2016-QNXZ-B-21 \& YBXM-2014-02) and the National Natural Science Foundation of China (Grants No. U1731103, 11573016 and 11503072). SF and K\'{E}G thank the Hungarian National Research, Development and Innovation Office (OTKA NN110333) for support. K\'{E}G was supported by the J\'{a}nos Bolyai Research Scholarship of the Hungarian Academy of Sciences. This work has made use of data from the European Space Agency (ESA)
mission {\it Gaia} (\url{https://www.cosmos.esa.int/gaia}), processed by the {\it Gaia} Data Processing and Analysis Consortium (DPAC, \url{https://www.cosmos.esa.int/web/gaia/dpac/consortium}). Funding for the DPAC has been provided by national institutions, in particular the institutions participating in the {\it Gaia} Multilateral Agreement.


\label{lastpage}

\end{document}